\documentclass{ws-ijmpc}

\begin{document}

\markboth{ Denis Horv\'ath, Zolt\'an Kuscsik}
{Self-modifying financial market network}

\catchline{}{}{}{}{}

\title{ Structurally dynamic spin market networks} 

\author{DENIS HORV\'ATH AND ZOLT\'AN KUSCSIK\footnote{Department of Theoretical Physics and Astrophysics, 
\v{S}af\'arik University, Park Angelinum 9, Ko\v{s}ice, Slovak Republic}}

\address{ Department of Theoretical Physics and Astrophysics, 
\v{S}af\'arik University, Park Angelinum 9,\\ 
 040 01 Ko\v{s}ice, Slovak Republic,
\footnote{horvath.denis@gmail.com}}

\maketitle

\begin{history}
\received{Day Month Year}
\revised{Day Month Year}
\end{history}

\begin{abstract}

The agent-based model of stock price dynamics on a directed evolving complex network 
is suggested and studied by direct simulation. The stationary regime is maintained 
as a result of the balance between the extremal dynamics, adaptivity of strategic variables and 
reconnection rules. The inherent structure of node agent "brain" 
is modeled by a recursive neural network with local and global inputs 
and feedback connections. For specific parametric combination the complex network 
displays small-world phenomenon combined with scale-free behavior. 
The identification of a local leader (network hub, agent whose strategies 
are frequently adapted by its neighbors) is carried out 
by repeated random walk process through network. The simulations show empirically 
relevant dynamics of price returns and volatility clustering. The additional emerging aspects 
of stylized market statistics are Zipfian distributions of fitness. 
 
\keywords{econophysics, complex networks, agent-based model} 

\end{abstract}

\ccode{PACS Nos.: 89.65.Gh, 89.75.Hc, 05.65.+b}

\section{Introduction}

The design of the decentralized systems of autonomous 
interacting agents with abilities to automatically 
devise societies so as to form modules that accommodate 
their behavior via social and economic norms in emergent 
ways is highly challenging. Agents in such groups plan and 
act in a way to increase their own utility. 

As a general theoretical framework for starting of the statistical 
consideration of interacting agent entities we take into account 
the Ising model. Being made simply by binary spin variables, the model 
is able to reproduce different complex phenomena in different areas of science like 
biology~\cite{Peng1992}, sociology~\cite{Schwetzer2000}
economy~\cite{Sznajd2003,Gordon2005,Kitsukawa2006} 
or informatics~\cite{Horiguchi2004}.

Looked at from the perspective of the economics this example has 
a great importance because it demonstrates 
that a basic interaction between the spins (agents) 
can bring non-trivial collective phenomena. 
The parallels between fluctuations in the economic 
and magnetic systems afford an application of spin models to the market statistics~\cite{Cont2000,Chowdhury1999,Silva2001}.  
The attempts~\cite{Ponzi2000,Bornholdt2001,Kaizoji2002,Badshah2005,Takaishi2005}
examine a context with the {\em minority game} theory~\cite{Challet1997}.
Basic terms borrowed from the magnetic systems have been 
built in to the spin agent concept: exchange interaction between spins and interaction 
with random field. The approaches based on the minority game assume 
competition of the short-range ferromagnetic and 
global magnetization terms that 
are crucial for modeling of bubbles and crashes. 
The 
justification for our present 
formulation comes from~\cite{Greco2006}.  
It assumes that each among $i=1,2, \ldots, L$ 
interacting 
traders (agents) owns her/his regular lattice site 
position $i$ and  a corresponding spin variable $S^{(t)}(i) \in \{-1,1\}$, 
where the upper index $(t)$ labels the market time. Each agent, 
has an attitude to place buy order $S^{(t)}(i)=1$ or to place sell order 
described by $S^{(t)}(i)=-1$. The variable $S^{(t)}(i)$ 
is updated by an asynchronous stochastic heat-bath dynamics expressed in 
terms of the {\em local effective field}. 

The collective effect of spins is characterized by the instant magnetization 
\begin{equation}
m^{(t)}= \frac{1}{L}\,\sum_{i=1}^L S^{(t)}(i)\,
\label{magne2}
\end{equation}
that is interpreted as measure 
price imbalance. Therefore 
the logarithmic price return can be 
written as~\cite{Farmer2002} 
\begin{equation}
\ln \left[ p^{(t+1)}/p^{(t)} \right] = m^{(t)}/\lambda\,,
\label{priceaccrel2}
\end{equation}
where $\lambda$ is the liquidity constant. The predominance of buy orders 
manifests itself throughout $m^{(t)}>0$. In that case the instant price 
$p^{(t+1)}$ exceeds $p^{(t)}$. In an analogous manner $m^{(t)}<0$ describes the fall 
of the stock price.

In the spin models the non-Gaussian contribution to distribution of magnetization 
is formed due to spin-spin interactions. The realistic models of social/economic networks can serve 
as topological substrate for spreading of the interaction effects. 
Our recent approach to spin market models~\cite{Horvath2006} has 
been formulated for network geometry. 
The attempt to postulate dynamics of
network that is coupled to the spin degrees of freedom is analogous to 
{\em structurally dynamic} {\em cellular automata}~\cite{Ilachin1987} 
in which the conventional cellular-automata rules are 
generalized to formalism where geometry and matter are dynamically coupled. 

Since our former attempt did not give satisfactory 
topological results, one of the aims of our present work 
is to propose richer and more reliable formulation.  

The main interests of our present research 
is focused on:
\begin{itemize}

\item[(1)]~An alternative protocol of the network reconnection assuming the 
           network segmented by local "authorities"(local leaders - hubs).

\item[(2)]~The suggestion of rules that support formation 
           of the "small world" and "scale-free" 
           paradigms for social networks. 

\item[(3)]~The including of the small-scale intra-agent 
           cognitive-like sensorial structures 
           sustained 
           by interactions within the 
           complex network of agents. 

\end{itemize}

The plan of the paper is as it follows. 
In the next section~\ref{net_section} we discuss the 
basic network properties. The overall dynamics are 
given in section~\ref{coev_section}. 
A more detailed definition of the model items appears 
in subsections~\ref{subsection_decision}-\ref{subsection_extremal}. 
In section~\ref{simul_section} we present statistical 
characteristics extracted from our simulation.

\section{The network topology}\label{net_section}

 Let's suppose that the market structure is determined 
 by the underlying complex network 
 defined by dynamical rules for active links between agents.
       Consider the directed network~(graph) 
       of labeled nodes 
       $\Gamma=  \{1,2,\,\ldots,\,L\}$,
       where node $i\in \Gamma$ attaches 
       via $N_{\rm out}$  
       directed  links to its neighbors $X_n(i) 
       \in \Gamma$, 
       $n\in I_{\rm out} \equiv 
       \{1,2,\ldots, N_{\rm out}\}$, 
       i.e. the graph is $N_{\rm out}$-regular.  
       Two outgoing links $X_1(i)=1+i \mbox{mod} L$,
       $\,X_2(i)=1+(L+i-2) \mbox{mod} L$ 
       of node $i\in \Gamma$ create 
       the bidirectional 
       cycle {\em static subgraph} 
       ($L$-gon).  The reconnection rules 
       are applied exceptionally to the 
       links $X^{(t)}_n(i)$, 
       $n\in  {\overline I}_{\rm out}$ $= 
       \{n_1; 3 \leq n_1 \leq N_{\rm out};\, n_1\in I_{\rm out} \}$.   
       The using of static module 
       ($\{1,2\}\subset I_{\rm out}$) 
       guarantees the 
       preservation of network 
       connectedness 
       at any stage $t$.  

\section{The formalism of co-evolutionary dynamics}\label{coev_section}

Formally, the stochastic 
co-evolutionary dynamics of agents can be described by the recursive formula 
\begin{equation}
{\overline \Pi}^{(t+1)} =  {\hat {\bf U}}(\,\overline{\Pi}^{(t)})
\label{Eqsto2}
\end{equation} 
including 
the composed configuration 
\begin{equation}
{\overline{\Pi}}^{(t)} 
\equiv  \left \{ \Pi^{(t)}(1), 
\Pi^{(t)}(2), \ldots, \Pi^{(t)}(L) \right \} 
\end{equation}
that consists of single-agent particulars  
\begin{eqnarray}
 \Pi^{(t)}(i) \equiv 
\left\{ 
\begin{array}{ll} 
\mbox{\small intranet 
intra-agent spins}   &   \Pi^{(t)}_{\rm ss}(i)   
 \equiv  \{  s^{(t)} (i,k)  \}_{ k \in \Gamma_{\rm intr} } 
\\
\mbox{\small strategic variables}  &   
\Pi^{(t)}_{\rm J}(i) =  
\{ J_{\rm intr}^{(t)}(i,k,q)\},\,\, 
\\ & \mbox{where}\,\, 
k, q \in \Gamma_{\rm intr}   \\    
\mbox{\small network links}        &  
\Pi^{(t)}_{\rm X}(i)   
\equiv  \{ X_{n}^{(t)}(i)\}\,,\, n \in I_{\rm out}\,. 
\\ 
\end{array}
\right.\,
\end{eqnarray}
The intranet of agent 
$i$ includes fully connected 
recurrent network consisting of nodes 
$\Gamma_{\rm intr}\equiv \{1,2,\ldots, N_{\rm intr}\}$ nodes 
[for more details see modification Eq.(\ref{Eq11})]. 
The nonlinear operator ${\bf {\widehat U}}$ 
entails the overall effect of the following 
single-agent operators 
\begin{eqnarray}
\begin{array}{llll}
\mbox{\small local field}  &  {\widehat U}_{\rm ss}(i)      &   
\mbox{\small acting on}    &   \Pi^{(t)}_{\rm ss}(i)  \\
\mbox{\small }   &  {\widehat U}_{\rm Ad}(i_{\rm a})        &   
\mbox{\small acting on}    &   \Pi^{(t)}_{\rm J}(i_{\rm a})       \\
\mbox{\small reconnection} &  {\widehat U}_{\rm Re}(i_{\rm r})        &  
\mbox{\small acting on}    &    \Pi^{(t)}_{\rm X}(i_{\rm r})                     \\
\mbox{\small extremal dynamics} & {\widehat U}_{\rm Ex}(i_{\rm minF}) &   
\mbox{\small acting on}    &   \Pi^{(t)}_{\rm ss}(i_{\rm minF})\,, 
                               \Pi^{(t)}_{\rm J}(i_{\rm minF})\\
\end{array}
\end{eqnarray}
where $i_{\rm minF}$ 
is defined by minimum of fitness [see Eq.(\ref{fitness1}) 
in further] 
\begin{equation}
F(i_{\rm minF}) = \min_{j\in \Gamma} F(j)\,.
\label{minF2}
\end{equation}

The pseudo-code corresponding to dynamics Eq.(\ref{Eqsto2}) 
is described in the next five steps:

\begin{itemize}

\item {\bf loop I} over the Monte Carlo steps per network node   
 
\hspace{1.2cm} {\bf loop II} over the network of agents

  \vspace{0.1cm}

  \hspace{3.3cm} 1. pick agent $i$ randomly
 
  \hspace{3.3cm} 2. perform the field 
                    and spin update 
                    ${\widehat U}_{\rm ss}(i)$ 
                 
  \hspace{3.6cm}  [see Eqs.(\ref{locfield2}) and (\ref{field2}) of 
                  subsection \ref{subsection_decision}]. 

  \hspace{3.3cm} 3. pick two agents randomly: 

  \hspace{3.7cm} $i_{\rm a}$ for the adaptivity rule and 

  \hspace{3.7cm} $i_{\rm r}$ for the reconnection rule.

  \hspace{3.3cm} 4. apply $\,{\widehat U}_{\rm Ad}(i_{\rm a})$ 

  \hspace{3.7cm} with probability 
                 $P_{\rm Ad}$~(see subsection~\ref{subsection_})  

  \hspace{3.3cm} 5. apply $\,{\widehat U}_{\rm Re}(i_{\rm r})$ 

  \hspace{3.7cm} with probability
                 $P_{\rm Re}$~(see subsection~\ref{subsection_reconnection})

  \vspace{0.2cm}

  \hspace{1.2cm} {\bf end} of {\bf loop II}

  \vspace{0.1cm}

  \hspace{1.2cm} apply the extremal 
                 dynamics ${\widehat U}_{\rm Ex}(i_{\rm minF})$ 
                 to agent $i_{\rm minF}$ 
                 having 

  \hspace{1.2cm} the smallest fitness within 
                 the entire system.  

  \hspace{0.4cm} {\bf end} of  {\bf loop I}

  \vspace{0.6cm}

\noindent After it the following steps are carried out:

\vspace{0.2cm}

\item  (i) \hspace{0.3cm}  store instant $m$ calculated 
 according Eq.(\ref{magne2}), 

\item (ii) \hspace{0.2cm} calculate the 
            {\em price return} 
            Eq.(\ref{priceaccrel2}) and volatility as $|m^{(t)}|$ 

\item (iii) \hspace{0.01cm} update 
            {\em fitness} 
            $F(i)$  
            for all agents 
            according 
            Eq.(\ref{fitness1}) 
            (see subsection~\ref{subsection_fitness4})\,.

\end{itemize}

\subsection{${\widehat U}_{\rm ss}$: 
            decision making via local 
            field,   
            interface between extranet 
            and intranet}\label{subsection_decision}

Through links $X_n(i)$ agent $i$ 
gains the game-relevant information 
about the external nearest world. For directed 
network topology we define the integral spin value   
\begin{equation}
 S_{\rm nn}(i)   
 \leftarrow  
 \frac{1}{N_{\rm out}}  \,   \sum_{n\in I_{\rm out}}  
  \,  S(X_n(i))\,. 
 \label{locfield2}
 \end{equation}

 The intra-agent idiosyncratic 
 structure is modeled in a abstract way that resembles the 
 modular organization 
 of the routing control unit~\cite{Horiguchi2004}. 
 The present variant 
 of the model goes essentially 
 beyond the elementary single-spin intra-agent 
 description. We have 
 looked in this direction by considering trader's states encoded by 
 Ising spins (neurons) coupled by the weighted links of 
 the {\em fully interconnected intranet}. 
 We suppose the agent's $i$ architecture consisting of 
 $N_{\rm intr}$ spins 
 $\{ s(i,k) \}_{k\in \Gamma_{\rm intr}}$, $s(i,k)\in \{-1,1 \}$. In 
 further, 
 we distinguish  between the inner 
 $s(\cdot,\cdot)$ 
 and extranet 
 $S(\cdot)$ symbols.  
 The effective local field $h_{\rm loc}(i,k)$ is suggested 
 in the form 
 \begin{equation}
 h_{\rm loc}(i,k)        
\leftarrow    
 h_{\rm stoch}(i,k)     +  
\frac{1}{N_{\rm intr}-1}   
\sum_{ 
\mbox{
\scriptsize   
\mbox{
 $\begin{array}{c}   
 q \neq k  
\\  
 q, k\in \Gamma_{\rm intr} 
\end{array}$ } }}^{N_{\rm intr}}  
J_{\rm intr}(i,k,q) s(i,q)\,, 
\label{field2}
\end{equation}
where $h_{\rm stoch}(i,k)$ is the Gaussian stochastic variable 
(i.e. the source of local mutations) and $J_{\rm intr}(i,k,q)$ is 
the $N_{\rm intr} \times N_{\rm intr}$ system 
of the intra-agent pair couplings (weights); 
the term $1/(N_{\rm intr}-1)$ 
is included for normalization reasons. 
Three inputs of intranet are supposed:
\begin{equation}
s(i,3)  \leftarrow   m\,,    \qquad  
s(i,4)  \leftarrow   S(i)\,, \qquad  
s(i,5)  \leftarrow   S_{\rm nn}(i) \,.  
\label{s345Eq}
\end{equation}
Two of the spins $s(i,k)$, $k=3,5$ are linked to outer complex network represented 
by $m$ and $S_{\rm nn}(i)$. The feedback to $s(i,4)$ is related to self-control.  
The state of $k$th unit of agent $i$ is recalculated according  
\begin{equation}
s(i,k)  \leftarrow {\rm sign}  
\left(h_{\rm loc}(i,k) 
\right)\,, 
\qquad \mbox{when}\qquad 
k\in {\overline \Gamma}_{\rm intr}=  \{ 1,2 \}   \cup  \{ 6,7,  \ldots, N_{\rm intr}  \}\,.  
\label{Eq11}
\end{equation}
The remark here is that recursive spin update of intranet 
is performed asynchronously. 
Two-neuron output of intranet has been considered: 
$k\in \{1,2\}$, [see Eq.(\ref{field2})]. 
The agent's sell buy order  
\begin{equation}
S(i) \leftarrow \frac{1}{2}\left[  s(i,1) 
+  s(i,2)  \right]\,
\label{S12Eq}
\end{equation}
can be identified by other agents. The factor $1/2$ 
normalizes the state space $\{-1,0,1 \}$ of $S(i)$. 
The $S=0$ state is interpreted as a {\em passive}~\cite{Takaishi2005}. 
The impact of local $S_{\rm nn}(i)$ and global $m$ 
in Eq.(\ref{s345Eq}) may be 
interpreted as Keynes' beauty contest~\cite{Keynes34} according 
which the stock market notably 
reflects the mass psychology. 

The agent's decision sell or buy is given by the procedure:

\hspace{2mm}

\noindent \underline{ {\bf procedure} ${\hat U}_{\rm ss}(i)$}

\noindent {\bf loop III} over $N_{\rm intr}$ repetitions 

\hspace{12mm} 1.  pick $i\in \Gamma$ randomly;

\hspace{12mm} 2. calculate $S_{\rm nn}(i)$ according Eq.(\ref{locfield2});

\hspace{12mm} 3. make settings Eq.(\ref{s345Eq});

\hspace{12mm} 4. choose randomly intranet spin $k \in {\overline\Gamma}_{\rm intr}$; 

\hspace{12mm} 5. calculate the local field at $k$ according 
                 Eq.(\ref{Eq11});

\hspace{16mm}  the stochastic field $h_{\rm stoch}(i,k)$ is calculated 
               for each node separately;

\hspace{12mm} 6. update $S(i)$ 
              according Eq.(\ref{S12Eq});

\noindent {\bf end} of {\bf loop} {\bf III}

\noindent The {\bf output} of the procedure 
${\hat U}_{\rm ss}$ is 
$S(i)$ variable which can be identified 
by other linked agents [see Eq.(\ref{locfield2})]. 

\subsection{Fitness $F(i)$ concept} 
\label{subsection_fitness4}

Biological species interact with each other in ways which 
either increase or decrease their 
fitness. The concept is generalizable to survival of strategies undergoing selection 
of the 
economic agents. The {\em extremal dynamics}~[see Eq.(\ref{minF2})] 
governing co-evolution is based on the knowledge of the local 
fitness $F(i)$  that expresses 
an ability/inability 
to survive in the competitive environment. 
The selection and survival according 
fitness has been used for spin market models~\cite{Ponzi2000}. 
The fitness-dependent link formation~\cite{Sole1996,Ergun2001,Zheng2003} 
is conceptually close to our present formulation. In our model, 
the local fitness is defined as 
the integral over the history of agent's 
gains and losses calculated from market situation and sell buy orders 
\begin{equation}
F^{(t+1)}(i)  =  F^{(t)}(i)  +  S^{(t)}(i) \left[ - c_0 m^{(t)} +  c_{\rm rand}\, {\rm N}^{(t)}(0,1) \, 
\right]\,.
\label{fitness1}
\end{equation}
The formulation is based on the combination of {\em minority game} profit and external 
influences. The relationship of individual spin $S$ and majority is quantified 
by the $- c_0 S^{(t)}(i) m^{(t)}$ term with $c_0>0$. The impact of exogenous 
news~\cite{Sornette2006} is included via random Gaussian term (white noise) 
$c_{\rm rand} N^{(t)}(0,1)$ that is common for all agents. 

\subsection{ Adaptivity procedure ${\widehat U}_{\rm Ad}(i_{\rm a})$, herding}
\label{subsection_}

      Many mathematical models
      in the social sciences assume that humans 
      can be described as "rational" entities.  
      Nevertheless, the most people 
      are  only partly rational and in fact  are
      emotional/irrational in their decisions. 
      This property 
       is incorporated into 
      the concepts of {\em bounded rationality}~\cite{Gigerenzer2002} and
      {\em herding} of agents. 
      As a concrete example can serve the 
      followers with their believe 
      that imitation 
      of given strategy owned by the social neighbors 
      (selected with care according to their fitness) would bring a future 
      benefit to them. The  assumes 
      the transfer of 
      information in format  
      $\Pi_{\rm J}$  along the edge of extranet. 
      The adaption of 
      $i_{\rm a}$ starts with the random  
      pick of $n_{\rm a} 
      \in I_{\rm out}$  
      that checks  
      a {\em prototype node} 
      $i_{\rm prot}= X_{n_{\rm a}}(i_{\rm a})$ 
      among the nearest neighbors. 
      It should be stressed that links 
      are placed by requesting fitness preferences, therefore, 
      the adaptation is random only in the sense that 
      it is not directed 
      to specific $n_{\rm a}$. 

      The adaption  of $\Pi_{\rm J}(i_{\rm a})$ to $\Pi_{\rm J}(i_{\rm prot})$  is described 
      by updates  
\begin{eqnarray}
J_{\rm intr}(i_{\rm a},k,q)  
&\leftarrow &    
J_{\rm intr}(i_{\rm a},k,q)  
(1-\eta)  +  
\eta    
J_{\rm intr}(i_{\rm prot},k,q)\,,
\label{eqadapt2}
\end{eqnarray} 
where plasticity parameter $\eta \in (0,1)$ expresses  
how quickly the follower $i_{\rm a}$ learns a strategy. 
The repeated application of ${\widehat U}_{\rm Ad}(i_{\rm a})$ yields 
entropy consumption and strategic uniformization of market 
that works against distinctive thinking. 

\subsection{ Reconnection rules ${\widehat U}_{\rm Re}(i_{\rm r})$, 
network dynamics}\label{subsection_reconnection} 

 Recently, the interest in complex networks 
 has been extended to the seeking 
 for the local rules governing  
 the  build-up of social and technological networks. 
 Several principles have been exploited 
 for this purpose. As an example we mention the 
 network that shows marks of age~\cite{ZhuWangZhu2003} 
 or inter-agent communication across 
 the net~\cite{Anghel2004,Zimmermann2004}. 
 The core of these methodologies rely 
 on the particular mechanism of 
 {\em preferential attachment} suggested
 by Barab\'asi and Albert~\cite{Albert1999}. 
 A principal distinction from mentioned work 
 is that our present study directs attention 
 on stationarity conditions and constant vertex number. 

 Let us turn attention to our former proposal of centralized 
 {\em single leader model}~\cite{Horvath2006}. 
 Intuitively, it is implausible
 that $L-1$ followers 
 can continuously identify the leader
 and its strategy can be adapted by them in the limit of very large $L$. 
 The reason for the revision of single leader picture is that 
 the larger the market is, a more demanding and time consuming is the technical analysis 
 of the follower that claims 
 to localize socially attainable 
 local leader. 
 Accordingly, we are interested 
 in a more general multi-leader stationary 
 society influenced by a varying group of 
 irregularly distributed {\em local leaders}.  

 This paper aims to suggest indirect method of generation 
 of {\em segmented market} which includes many competing local leaders. 
 The network is reconstructed indirectly via the knowledge 
 extracted from the {\em random walk process on the 
 net}~\cite{Bray1988,Almaas2003}. 
 The walk is carried out by the "assistant/informant" agent 
 several times emitted from the same source node $i_{\rm r}$. 
 The informant behaves like the 
 Web surfer reading pages, 
 jumping from one to another by 
 clicking randomly on web links. 
 To be more precise, we define the 
 notion of {\em repeated random walk} as a set 
 of linked nodes  
 ${{\bf RRW}}(i_{\rm r}, N_{\rm rep}, N_{\rm depth} )\subset \Gamma$. 
 The set is obtained by performing $N_{\rm path}$ 
 steps that are $N_{\rm rep}$ times repeated from the origin 
 $i_{\rm r}$ occupied 
 by the agent seeking for a nearby local leader. 
 The fitness ranked 
 by walker - informant allows to propose preferential 
 attachments that are expected to yield formation of 
 {\em segmented market} with impediments 
 to the free flow of information. 
 The procedure of  {\em reconnection} to a
 local leader denoted as ${\widehat U}_{\rm Re}(i_{\rm r})$  
 is build upon the sub-procedure of edge {\rm pruning} ${\widehat U}_{{\rm Re}_{\rm w}}(i_{\rm r},i_{\rm B})$ 
 that calls the {\em best connection proposal} 
 ${\widehat U}_{{\rm Re}_{\rm B}}(i_{\rm r})$. 
 The sub-procedures are specified in bellow:   

\vspace{3mm}

\noindent \underline{{\bf procedure} 
          ${\widehat U}_{{\rm Re}_{\rm B}}(i_{\rm r})=i_{\rm B}$} 

\vspace{2mm}

({\bf 1}). \hspace{0.4cm} {\bf loop IV} 
                over $N_{\rm rep}$ 
                repetitions
 
\hspace{2.0cm}  that start from the initial 
                condition 
                $i_1\equiv i_{\rm r}$ 

\hspace{2cm}  {\bf loop} {\bf V} the execution 
               of 
               $N_{\rm depth}$ 
               iterations 
\begin{equation}
\qquad  i_{l+1}=X_{n_l}(i_l)\,, 
\qquad  l=1,\ldots, 
        N_{\rm depth}\,\,\, , 
\end{equation}

\hspace{2cm}    for random choices links 
                $n_l \in  I_{\rm out}$  

\hspace{2cm}    {\bf end} of {\bf loop} {\bf V}\,; 

\hspace{1.0cm}  {\bf end} of {\bf loop} {\bf IV}\,;

({\bf 2}). \hspace{0.4cm}   the {\em comparison} of 
                    $N_{\rm depth} \times 
                    N_{\rm rep}$ 
                    values of 
                    $F(i_l)$ 

({\bf 3}). \hspace{0.4cm}   and {\em localization} 
                    of the agent 
                    $i_{\rm B}$ according 

\begin{equation}
F(i_{\rm B})=   \max_{ i_l\in {\bf RRW}} F(i_l)
\end{equation}

\hspace{0.4cm}   via the set  
                 ${\bf RRW}(i_{\rm r}, N_{\rm rep}; 
                 N_{\rm depth})$ bounded 
                 by the radius $N_{\rm depth}$

\hspace{0.4cm} formed by nodes 
               $i_1, i_2, \ldots,  i_{N_{\rm depth}}$ visited according 
    
\hspace{0.4cm} loops {\bf IV}, {\bf V}.  
 
\noindent The {\bf output} of 
          ${\widehat U}_{{\rm Re}_{\rm B}}(i_{\rm r})$ 
          is agent  
          $i_{\rm B}\in \Gamma$ that is the
          {\em candidate} 
          for future connection 
          from node $i_{\rm r}$. 

 \vspace{8mm} 

 \noindent The pruning is performed according:

 \noindent \underline{{\bf procedure} 
 ${\widehat U}_{{\rm Re}_{\rm w}} (i_{\rm r}, i_{\rm B})
 =  {\widehat U}_{{\rm Re}_{\rm w}} (i_{\rm r},  {\widehat U}_{{\rm Re}_{\rm B}} (i_{\rm r}))=n_{\rm w}$}

 \hspace{0.4cm} {\bf loop VI} over the 
                $N_{\rm out}$ 
                nearest  
                neighbors. 
                Determine the "weakest"   

  \hspace{2.1cm} ({\em worst}) connection 
                 $n_{\rm w}$ 
                  with the smallest 
                  fitness 

\begin{equation}
F(X_{n_{\rm w}}(i_{\rm r})) =  
\min_{n \in {\overline I}_{\rm out}} \, 
F(X_{n}(i_{\rm r})) 
\end{equation}

 \hspace{2.1cm} within 
                the actual existing 
                connections 

 \hspace{0.4cm} {\bf end} of {\bf loop VI}\,. 

\noindent The 
           {\bf output} of  
          ${\widehat U}_{{\rm Re}_{\rm B}}(i_{\rm r})$  
          is the 
          index $n_{\rm w}\in {\overline I}_{\rm out}$.
   
  \vspace{0.4cm}

  \noindent Finally, if 
  $F(X_{n_{\rm w}}(i_{\rm r}))<F(i_{\rm B})$ 
  the {\bf output} of ${\widehat U}_{\rm Re}(i_{\rm r})$ 
  is the update 
  \begin{equation}
   X_{n_{\rm w}}(i_{\rm r}) \leftarrow i_{\rm B}
  \end{equation}
  conditioned by the 
  requirement that no 
  multiple 
  connections between 
  $i_{\rm r}$ 
  and $i_{\rm B}$ are established. 
  It also forbids self-connection 
  loops 
  ($X_n(i_{\rm r})=i_{\rm r}$). 

 \begin{figure}
 \epsfig{figure=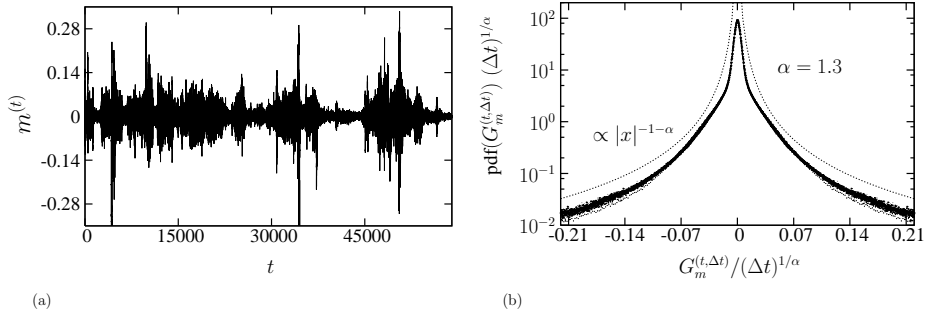,width=12.2cm}
 \caption{\small 
 (a) The time evolution of the log-price returns defined as magnetization
 $m^{(t)}$. Part (b) shows the scaling properties of the pdf of cumulated returns 
 $G_m^{(t,\Delta t)} = m^{(t+1)}+m^{(t+2)}+ \ldots+ m^{(t+\Delta t)}$.   
 Parameter $\alpha=1.3$ 
 determined 
 at first from the fit well 
 coincides with 
 the scaling collapse of eight distributions 
 obtained for trading time separations 
 $\Delta t=1, 2, \ldots, 8.$).}
 \label{Fig1}
 \end{figure}
\begin{figure}
\epsfig{figure=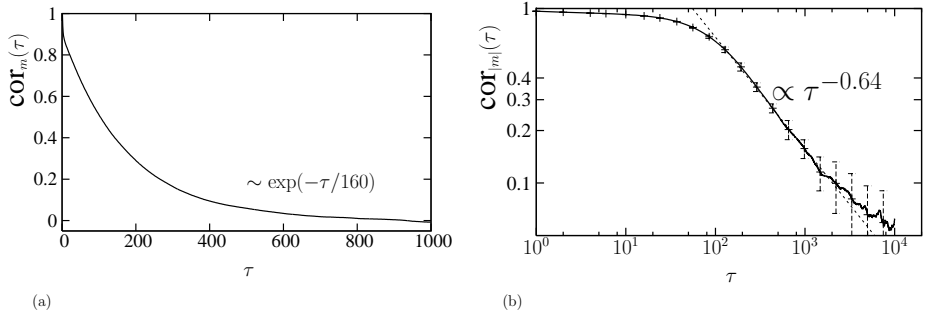,width=12.2cm}
\caption{\small 
(a)~The autocorrelation function of the 
logarithmic price return $m^{(t)}$. 
(b)~The indication of the power-law long-time 
 regime of the autocorrelation function of log-price volatility 
 $|m^{(t)}|$.} 
\label{Fig2}
\end{figure}
\begin{figure}
\epsfig{figure=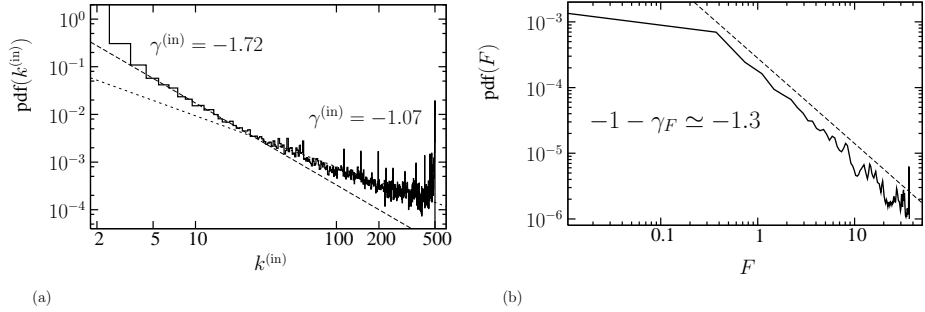,width=12.2cm}
\caption{\small 
(a)~The power-law pdf($k^{(in)}$) with 
    local effective exponents $\gamma^{\rm (in)}
=- 1.07$~(when $k^{\rm (in)}\geq 20$), and 
 $\gamma^{\rm (in)}=- 1.07$ [for 
  $k^{\rm (in)}\in (5,20)$].  
(b)~The Zipf's $\gamma_F=0.3$ 
 law of fitness.}
\label{Fig3}
\end{figure}
\begin{figure}
\epsfig{figure=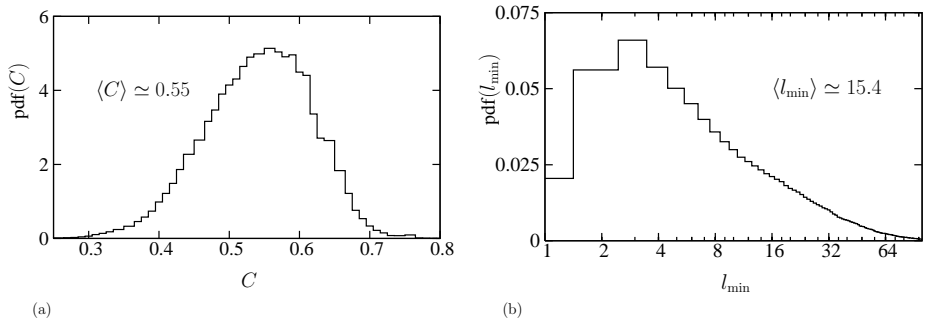,width=12.2cm}
\caption{\small 
(a)~The pdf of clustering 
coefficients. In 
part (b)~the pdf of the minimum path ways 
between randomly selected pair of nodes is plotted.}
\label{Fig4}
\end{figure}
 \begin{figure}
 \epsfig{figure=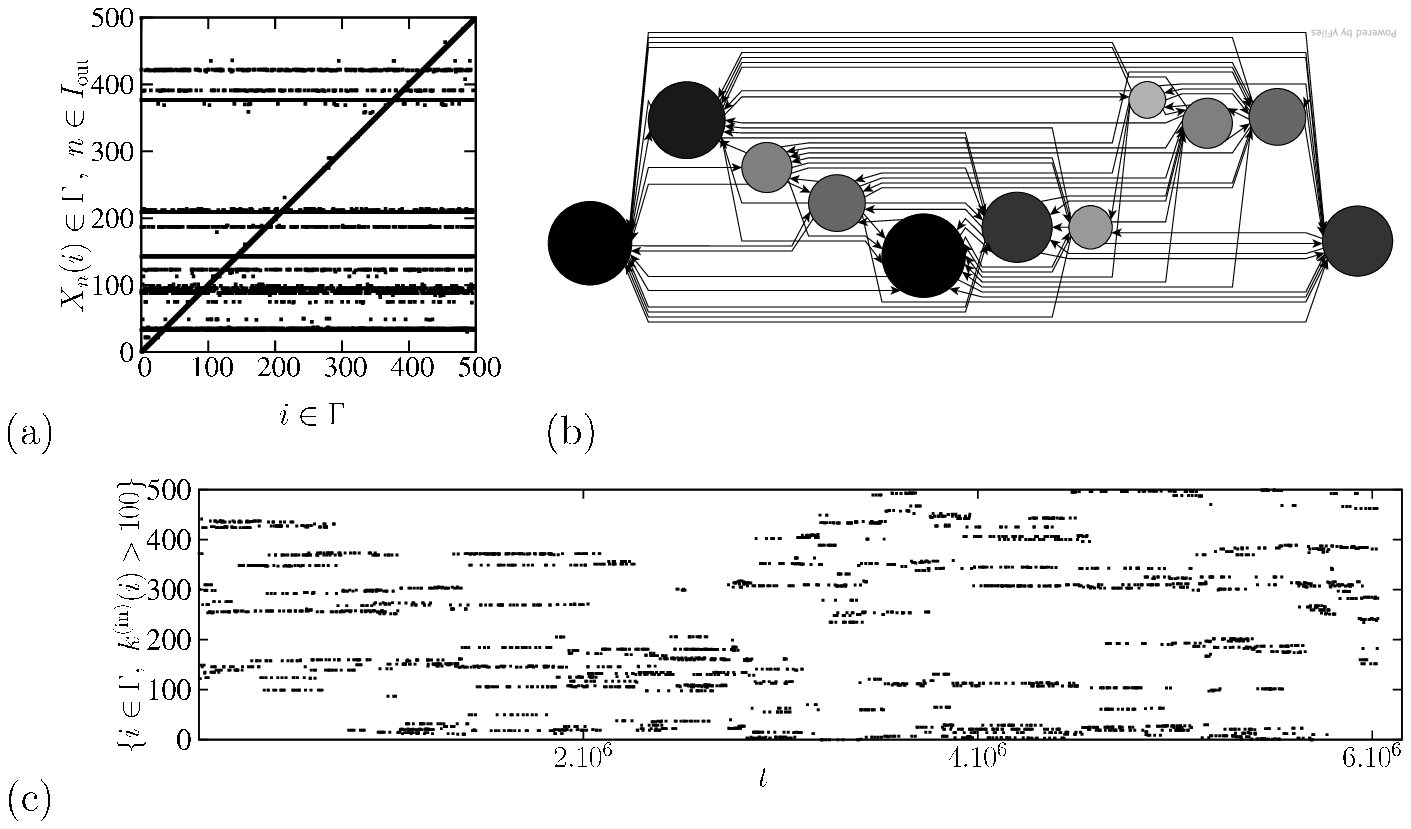,width=12.2cm}
 \caption{\small (a) 
  The snapshot of adjacency $L \times L$  matrix 
  (the point denotes the connection). 
  The cyclic L-gon maps onto the matrix diagonal. (b)~The subgraph 
  $\{i\in \Gamma, k^{\rm (in))}>50 \}$ (the larger circle belongs to larger $k^{\rm (in)}$.
  (c)~The epochs of topology monitored by means of 
  selection of highly preferred $\{i, i\in \Gamma,\, k^{(in)}(i)>100\}\subset \Gamma$.
  The members of such group may be vaguely identified as local leaders. }
 \label{Fig5}
 \end{figure}


\subsection{Extremal dynamics  ${\widehat U}_{\rm Ex}(i_{\rm minF})$, 
 entry and exit of market}
\label{subsection_extremal}

  The extremal dynamics~\cite{Bak1993,Flyvbjerg1993,Pienegonda2003} when interpreted 
  as a bankruptcy of firm 
  or exit of strategy from market 
  could be considered to be the principal mechanism of the economic {\em co-evolution}. 
  The idea of Bak-Sneppen model can be easily converted to 
  strategic variables.  In that case the least fit strategy developed by 
  $i_{\rm minF}$ is replaced by a random candidate from 
  certain limited strategic space. No reconnections are assumed within the procedure  
  ${\widehat U}_{\rm Ex}$ itself. This simplification is justified 
  by the assumption that low fitness agents are rarely attached 
  [i.e. $k^{\rm (in)}(i_{\rm minF})$ is relatively small]. This reluctance against 
  attachment  follows from the preferences incorporated into 
  ${\widehat U}_{\rm Re}(i_{\rm r})$. The 
  extremal event in the present formulation means 
  that instant value of the strategic 
  variable is immediately 
  replaced by Gaussian distributed random number 
  ${\rm N}(0,\sigma_{\{\ldots\}})$ 
  of dispersion 
  $\sigma_{\{\ldots\}}$. 
  The new strategy enters thought 
  \begin{eqnarray}
   J_{\rm intr}(i_{\rm minF},k,q)   &  
   \leftarrow & {\rm N}(0,\sigma_{J_{\rm intr}})\,, 
   \\
   h_{\rm stoch}(i_{\rm minF},k,q)  &  
   \leftarrow &  {\rm N}(0,\sigma_{h_{\rm stoch}})\,, 
   \nonumber 
   \\
   F(i_{\rm minF}) &  \leftarrow   &  
   {\rm N}(0,\sigma_{\rm F})\,. 
   \nonumber 
   \end{eqnarray}
   These updates necessitate the fixing of three independent dispersion parameters 
   $   \sigma_{J_{\rm intr}}\,,\,     
   \sigma_{\rm F}\,,\,   \sigma_{{\rm h}_{\rm stoch}}$.

  \begin{table}[ht]
  \tbl{List of 13 numerical parameters for which 
   the statistics is presented.}
  {\begin{tabular}{@{}|l|ll|l@{}} 
   \hline
   \mbox{number of}               &    \mbox{extranet nodes}  &   $L=500$  \,\\
 & \mbox{node outputs}            &     $N_{\rm out}=10$\,         \\
 & \mbox{random search steps}     &     $N_{\rm depth}=6$\,         \\
 & \mbox{repeated net searches}   &     $N_{{\rm rep}}=6$          \\
 & \mbox{intranet nodes}        &     $N_{{\rm intr}}=8$          \\
  \hline
 \mbox{dispersion of}     &     $J_{\rm intr}$      &   $\sigma_{J_{\rm intr}}=1$       \\
                          &     $h_{\rm stoch}$     &   $\sigma_{h_{\rm stoch}}=0.001$   \\
 &  $F$                   &     $\sigma_{\rm F}=0.1 $   \\                               
  \hline 
  \mbox{probability of}      &    \mbox{reconnection} &   $P_{\rm Re}=0.01$ \\
  & \mbox{adaptive move}     &    $P_{\rm Ad}=0.2$   \\
  \hline
  \mbox{adaptivity}          &    \mbox{parameter}&   $\eta=0.1$         \\
  \hline 
  \mbox{fitness}             &    \mbox{minority game}&            
  $c_0=1$     \\
                             &    \mbox{news}&    
  $c_{\rm ran}=0$ except Tab.\ref{Tab2}   \\
   \hline
  \end{tabular}
  \label{Tab1}}
  \end{table}

  \begin{table}[ht]
  \tbl{
    The study of unexpected information about 
    economic performance or political situations 
    incorporated into fitness. 
    The comparison of averages 
    corresponding to two combinations of 
    parameters 
    $c_0$, 
    $c_{\rm rand}$ of local  
    fitness Eq.(\ref{fitness1}). 
    The measure 
    $P_{m,4}$ 
    denotes the probability that 
    $m^{(t)}$ does not alter sign during the four 
    subsequent steps $m^{(t)}$, $m^{(t+1)}$, $m^{(t+2)}$, $m^{(t+3)}$ 
    (four times bearish or four times bullish stock).  
    The table shows 
    that risk aversion accompanied 
    by the probable passive 
    states with reduced 
    $\langle |m^{(t)}|\rangle$ 
    can issue from the high exogenous influence. 
    The calculation of 
    $P_{m,4}$ indicates that news can 
    break bearish (bullish) sequences. 
    The population of 
    local leaders quantified by 
    exceptionally connected nodes 
    characterized by probability 
    $P_{k>100} 
    \equiv \sum_{ k^{\rm (in)}>100 } \mbox{pdf}(k^{\rm (in)})$.}
   {\begin{tabular}{@{}|ll|ll|lll|@{}} 
   \hline
   $c_0 $            &                       
   $c_{\rm rand}$    &                       
   $\langle |m^{(t)}|\rangle $      &        
   $ P_{m,4}                 $      &        
   $ P_{k>100}               $      &        
   $\langle C \rangle $             &        
   $\langle l_{\rm min} \rangle$             
    \\
   \hline   
    $1$ &    $0.00$    &  $ 0.0138  $ &  $0.226$ &    $0.02$  &  $0.55$  & $15.2$    \\
    $1$ &    $0.05$    &  $ 0.0082  $ &  $0.165$ &    $0.02$  &  $0.56$  & $18.4$    \\
    $1$ &    $0.10$    &  $ 0.0029  $ &  $0.091$ &    $0.02$  &  $0.57$  & $15.2$    \\
  \hline 
  \end{tabular}
  \label{Tab2}}
  \end{table}

\section{Simulation results}\label{simul_section}

\subsection{Selection of parameters}

 The statistics as presented 
 in the next subsection is determined by 
 the choice of 13 free parameters. Their adjustment is a nontrivial task as 
 whenever in the modeling of agency. 
 To attain at least qualitative agreement 
 with current statistical 
 concepts in finance~\cite{Mantegna1999}, 
 the values (see Table~\ref{Tab1}) have been suggested by the optimization in the parametric space. 
 A particular requirement is to keep the spin dynamics much 
 faster than . An additional 
 constraint aims at attaining vicinity 
 of a {\em critical regime}, where the power-law distributions 
 are obeyed~\cite{Sole1996} that are much 
 studied topics inside Econophysics.  Various 
 mechanisms of the generation 
 of power-laws have 
 been recently summarized 
 in review article~\cite{Newman2006}.  

 \subsection{Distributions and averages}

 Starting from random initial condition, we let 
 the system evolve at least $5. 10^4$ Monte Carlo steps per node. 
 The data 
 for calculation of averages of 
 interest have been collected from the next $\sim 10^7$ Monte Carlo steps. 
 Their validity is verified for several independent trials. 
       The comparison of selected averages corresponding 
       to different 
       fitness is given by Tab.\ref{Tab2}. 
       The table indicates 
       that external news yield more careful 
       behavior of agents and preference 
       of  passive states $S=0$ that anomalously sharpen the central part 
       of probability density function (pdf) of $m$ denoted as 
       $\mbox{pdf}(m)$. For given example, 
       the impact of 
       $c_{\rm ran}$ on the topology is marginal. 
        
       In further we concentrate on the case $c_{\rm rand}=0$.  
       The statistical treatment~(see Fig.\ref{Fig1}) 
       leads to the fat tailed pdf($m$)
       that have been fitted in particular by the power law asymptotics $|m|^{-1-\alpha}$ of the realistic 
       effective exponent $\alpha\simeq 1.3$ \cite{Mantegna1999}.  
       The dynamical features 
       of the price dynamics are highlighted 
       by the autocorrelation 
       functions in Fig.\ref{Fig2}. The clustering 
       of the volatility $|m^{(t)}|$ 
       of the log-price returns 
       is observed. 
       In that case the power-law indicates the occurrence 
       of the long time memory. Despite of the fact that exponent $-0.64$ 
       does not strictly reproduce empirical findings $-0.34$ \cite{Stanley2000} and  $-0.2$~\cite{Zawadowski2004}, 
       the recovery of log-slope from $(-1,0)$ range is quite encouraging 
       for the perspectives of the model.  
       The situation is even more interesting 
       since the power-law pdf($F$) [see Fig.\ref{Fig3}(b)] histogramed for $F>0$ can be interpreted 
       as Zipf's $\gamma_{\rm F}=0.3$ law for 
       survival abilities of strategies. 
  
       Let us turn attention to the issues of network 
       statistics depicted in 
       Figs.\ref{Fig3}(a),\ref{Fig4} and \ref{Fig5}. 
       The {\em node degree} $k^{(\rm in)}(j)   =  
       \sum_{i\in \Gamma} $ 
        $ \sum_{n\in I_{\rm out}} 
       \delta_{j,X_n(i)}$ 
       accounts for incoming 
       links of node $j$. 
       The stationary regime 
       generates sequence of networks with broad-scale pdf of node degrees. 
       By fitting of pdf$(k^{(\rm in)})$ 
       we have identified 
       two partial 
       effective 
       exponents 
       $\gamma^{\rm in} \simeq -1.07$ 
       [for $k^{({\rm in})} \in (5,20)$]\,, 
       and $-1.72$ 
       (as $k^{({\rm in})} \leq 20$) 
       from the law 
       $\mbox{pdf}(k^{\rm in})\sim 
       [k^{({\rm in})}]^{\gamma^{\rm in}}$.  
       Several real exponents 
       are provided here 
       for illustrative purposes.  
       The value $1.81$~\cite{Ebel2002} 
       corresponds to the collection 
       of e-mail addresses. 
       The exponent $-1.2$ 
       belongs to the coauthorship 
       network~\cite{Newman2001}. 

In further the organization of the network is done using 
{\em clustering coefficient} 
$C(i)$~\cite{DorogMendes2004} that is 
a local measure of interrelatedness of triplets or   
{\em social} {\em transitivity}~\cite{Ebel2003}.  
For directed network the tendency can be measured by 
$C(i) = e(i)/ ( N_{\rm out} (N_{\rm out}-1)) $,  
where $e(i)  \equiv $          $\sum_{n_1,n_2,n_3=1}^{N_{\rm out} 
 \times  N_{\rm out} \times N_{\rm out}} $ $
       \delta_{X_{n_1}(X_{n_2}(i)),X_{n_3}(i)}$ 
       stands for the number of the links between neighbors 
       $X_{n_1}(X_{n_2}(i))$ and $X_{n_3}(i)$  
       attained from $i$. The maximum number of links $N_{\rm out} 
       ( N_{\rm out} -1 )$ normalizes the expression for $C(i)$.  
       The object of interest is 
       the mean $\langle C \rangle $.
       As usual, it is meaningful 
       to compare the mean clustering coefficients  
       of two distinct network reconnection modes. 
       For the network 
       with randomized~${\overline I}_{\rm out}$ links 
       we obtained  
       $\langle C_{\rm rand} \rangle  \simeq 0.02$,  
       while $\langle C \rangle \simeq 0.55$ ($\langle C \rangle /\langle C_{\rm rand} \rangle \simeq 27.5$)
       reached by ${\hat U}_{\rm Re}$ is much higher and thus empirically more 
       relevant~\cite{Ebel2003}.
       We have computed the average of the minimum path way 
       for the partially random net 
       $\langle l_{\rm min, rand} 
       \rangle \simeq 2.9$~(with the circular subgraph untouched). 
       The action of ${\widehat U}_{\rm Re}(i_{\rm r})$ provides 
       $\langle l_{\rm min} \rangle \simeq 15.4$ 
       however much smaller maximum $l_{\rm min,max}=3$ of 
       $\mbox{pdf} (l_{\rm min})$. The combination of above topological 
       attributes supports both small-world and scale-free concepts of network statistics. 
       Fig.\ref{Fig5} indicates that ${\bf RRW}(.,.,.)$ invokes 
       generation of modular topology of 
       several local densely connected leaders.

\section{Conclusions}

   In this paper we reveal some interesting properties of the model of spin dynamics on a complex 
   evolving network. We demonstrated that our model is relevant for the description of stock market
   statistics.  The stationary regime is formed due to balance between information entropy inflow produced 
   by the extremal dynamics and stochastic outer sources compensated by the entropy outflow caused 
   by the adaptive moves. 

   We are conscious that minority game term $-c_0 S^{(t)}(i)  m^{(t)}$ of fitness only roughly describes the financial 
   profits at the real stock markets. In further research we plan the comparison with $\$$-game model~\cite{Andersen2003} 
   that represents a step closer to stock market reality.  Additional improvements of fitness could be done 
   by considering the impact of regulatory and non regulatory halts and delays. 

   The most emergent aspects of agent systems near the criticality are the 
   power-law distributions 
   of the topological and spin characteristics that occcur as a consequence of self-organization processes. 
   The necessary remark in this context is that exponents of such dependencies are non-universal, 
   i.e. they can vary from one set of free parameters to another.
   
   The adjustment of parameters can be formulated as multi-objective optimization computationally demanding task. 
   Due to its comprehensive features, the problem is planned to be discussed in future works. 

   The site http://158.197.33.91/\%7Ehorvath/selfstructured\_net/ provides our simulation C++ code and 
   supplementary materials.  
 
 \vspace{2mm}

 The authors would like to thank for financial 
 support through grants VEGA 1/2009/05, VEGA 1/4021/07, 
 APVT-51-052702, APVV-LPP-0030-06.

\end{document}